\documentclass[twocolumn,preprintnumbers,amsmath,assume,nofootinbib,superscriptaddress,aps]{revtex4-1}
\usepackage{graphicx}
\usepackage{dcolumn}
\usepackage{bm}
\usepackage{comment}
\usepackage{natbib}
\usepackage{wrapfig} 
\begin{document}
\title{Interaction-Driven Asymmetry in the Breakdown of the $\nu$ = 1 Quantum Hall State}

\author{Hoai Anh Ho}

\altaffiliation[Also at ]{Physics Department, Wayne State University.}
\author{Jian Huang}
\email{jianhuang@wayne.edu}
\affiliation{%
Department of Physics and Astronomy, Wayne State University, Detroit 48201, Michigan, USA\\
}%
\author{L. N. Pfeiffer}
\author{K. W. West}
\affiliation{
Department of Electrical Engineering, Princeton University, Princeton 08544, New Jersey, USA \\
%
}


\date{\today}

\newpage

\begin{abstract}
\textbf{
We report real-time detection of longitudinal and transverse transport responses across distinct frequency bands in a ferromagnetic filling factor $\nu$ = 1 integer quantum Hall state. By tuning $\nu$, we simultaneously access the evolution of the screening environment and bulk excitation structure. The resulting asymmetric breakdown, for $\nu>1$ and $\nu<1$, reveals that interaction effects, rather than a single-particle band picture, dominate the transport instability. Our findings highlight the indispensability of electron-electron interactions even in integer quantum Hall phases, suggesting that distinct many-body entanglement structures underlie both integer and fractional topological phases.
}
\end{abstract} 

\pacs{Valid PacS appear here}
\maketitle

\section{INTRODUCTION}

The integer quantum Hall effect (IQHE)~\cite{klitzing1980new} is conventionally regarded as a paradigmatic band-topological phase: a filled Landau level (LL) with Chern number $C=1$ yields a quantized Hall conductance $\sigma_{xy} = e^2/h$ protected against disorder~\cite{TKNN}. In this single-particle formulation, the essential ingredients are Landau-quantized wavefunctions and band topology, while electron-electron ($e$-$e$) interactions are presumed irrelevant or perturbative~\cite{Prange1990QHEBook}. By contrast, fractional quantum Hall (FQH) phases~\cite{FQHE} are understood as inherently many-body states, stabilized by interactions and characterized by fractionalized excitations and topological order~\cite{laughlin1983anomalous,wen1990topological}.

This dichotomy, however, is incomplete. At filling factor $\nu=1$, the ground state of a two-dimensional (2D) electron or hole gas is not a trivial band insulator but a ferromagnet stabilized by exchange interactions~\cite{girvin1999quantum}, with skyrmions as the lowest-energy charged excitations~\cite{sondhi1993skyrmions,barrett1995optically,schmeller1995evidence}. Thus, $e$-$e$ interaction is essential even in the simplest integer state. More generally, the entanglement structure of many-body wavefunctions~\cite{Li2008Entanglement,kitaev2006topological}, rather than the topology of a single particle alone, may govern the robustness of integer phases.

Breakdown phenomena provide a particularly sensitive probe of these entanglement structures~\cite{FonteinCorbino1988,komiyama1992inter,nachtwei1999breakdown,hoai2024topological}. When a quantized Hall state is driven beyond its stability limit, dissipation appears as the collective state destabilizes, revealing the interplay of interactions, disorders, excitations, and topology. However, the interpretation of previous results remains ambiguous, as the dominant localization mechanism can shift depending on the balance between disorder and interactions. Experimental conditions such as temperatures and excitation and detection techniques are also important. A direct, experimentally grounded correlation effect is essential for understanding topological protection, but such characteristics have been challenging to establish~\cite{roulleau2008direct}.

Here, we study the breakdown of the $\nu=1$ IQHE in a Corbino geometry~\cite{halperin1982quantized,FonteinCorbino1988}, which allows simultaneous and independent detection of longitudinal and transverse transport responses. Using a dual-frequency scheme, we achieve real-time, frequency-resolved separation of orthogonal channels, eliminating edge contributions and contact artifacts. We find many-body-driven dielectric responses characterized by a pronounced asymmetry in breakdown behaviors between $\nu < 1$ and $\nu > 1$, a feature that cannot be reconciled with a single-particle picture. Our results demonstrate that many-body entanglement plays a decisive role in the stability of integer topological phases, narrowing the conceptual distinction between integer and fractional quantum Hall effects.

\section{METHODS}

We employ two-dimensional hole systems hosted in ultra-high-purity (100) GaAs/AlGaAs quantum well structures, patterned into Corbino geometry~\cite{halperin1982quantized, FonteinCorbino1988}. The carrier density is $p \approx 4 \times 10^{10}$ cm$^{-2}$, with mobility $\mu \approx 2.5 \times 10^6$ cm$^2$/Vs. The effective mass $m^* \approx 0.45m_0$ yields a large ratio of Coulomb to cyclotron energy, $E_{\mathrm{Coulomb}}/\hbar \omega_c \sim 60$ in a perpendicular magnetic field $B=1$ T, ensuring that the $\nu=1$ state is strongly interaction-dominated. $h$ is the Planck constant and $\omega_c$ is the cyclotron frequency. The disorder is minimal: potential fluctuations have a correlation length $\lambda \gg l_B \sim 30$ nm, consistent with long-range scattering due to remote impurities in ultra-high-mobility GaAs/AlGaAs systems~\cite{coleridge1989low,sarma1985single}, suppressing localization effects that could obscure intrinsic many-body physics.

\begin{figure}[h]
\vspace{0pt}
\includegraphics[width=0.49\textwidth]{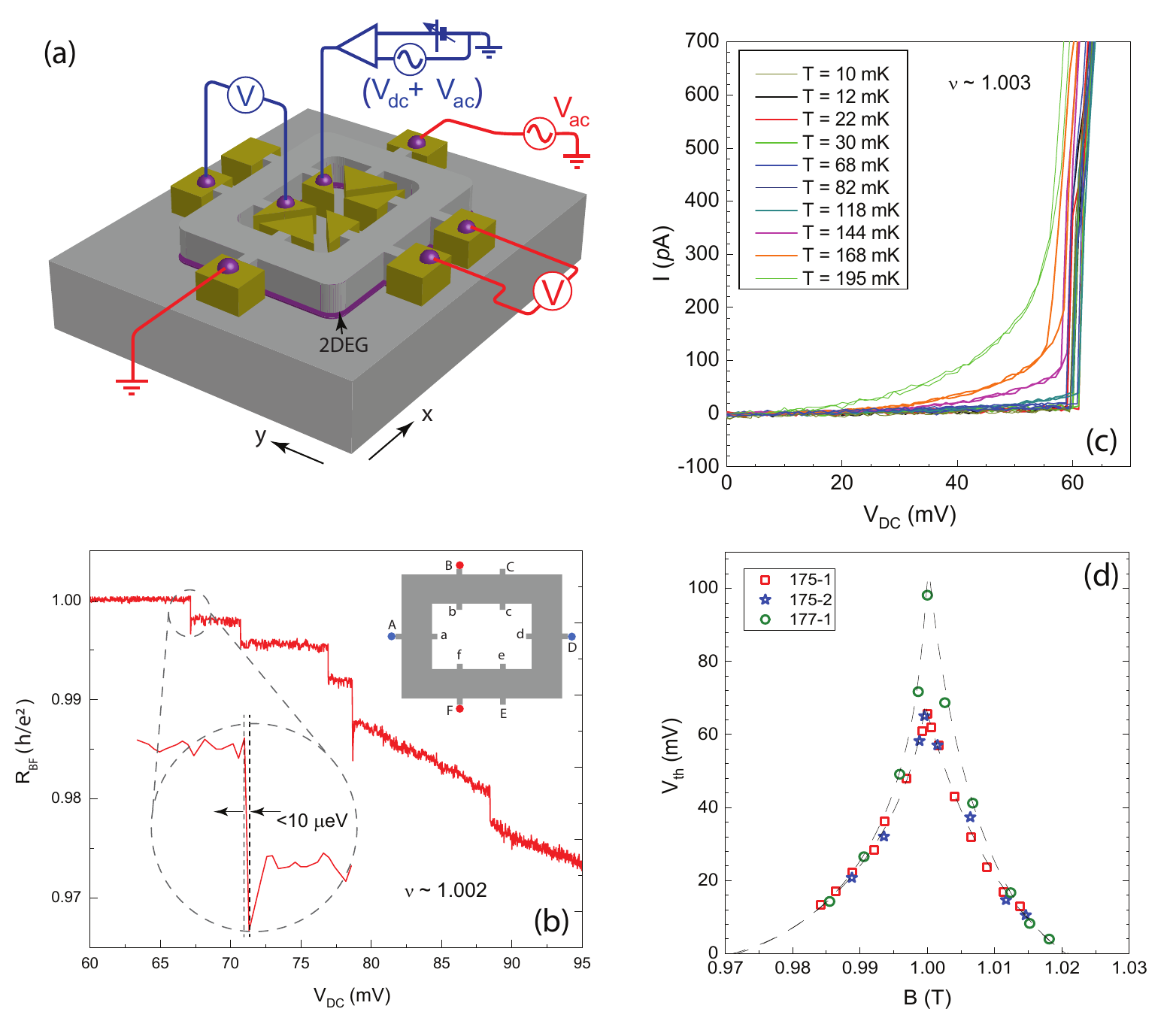} 
\vspace{-15pt}
\caption{\label{fig:Fig1}
(a) Schematic of the device and measurement setup, showing independent longitudinal and transverse detections. (b) Definition of breakdown: $R_{xy}$ (in $h/e^2$) vs. $V_{DC}$ at $\nu\sim1.002$ and T=15 mK, showing sharp, step-like deviations from quantization. Inset: sourcing contacts (blue dots) and measurement contacts (red dots) in the Corbino geometry. (c) T-dependence: Current I vs. $V_{DC}$ for $\nu\sim$1.003 at various T from 10 to 200 mK. (d) Critical breakdown voltage $V_{th}$ as a function of filling deviation $\Delta\nu=\nu-1$ obtained for three different samples (samples 175-1 and -2 are from the same wafer, 177-1 from a different wafer), all showing an exponential dependence (guide to the eye) asymmetric about $\nu=1$.}
\end{figure}

The additional electrical contacts, as illustrated in ~Fig.1(a), permit orthogonal probing of longitudinal and transverse dissipation channels, with independent electronics designed to handle the large anisotropy. Transport measurements employ a dual-frequency lock-in technique. A transverse electric field with both DC and AC components (46 to 50 Hz) at the $\sim 10$ $\mu e$V  scale is applied, enabling differential conductance measurements with sub-femtoampere sensitivity, facilitated by an electrometer within the triaxial wiring arrangement~\cite{hoai2024topological}. Simultaneously, a longitudinal AC excitation at a distinct frequency (of 7~Hz) is applied, enabling clean frequency-domain separation of dissipation channels. This ensures breakdown thresholds are determined with $\sim 10$ $\mu e$V precision, unaffected by mixing.

\section{RESULTS}
Figure 1(b) defines the breakdown, marked by sharp step-like deviations of the Hall resistance from $h/e^2$ quantization. For $\nu = 1.002$, $R_{xy}$ exhibits abrupt jumps above a threshold bias $V_{\mathrm{DC}} \approx 64$ mV. Such step-like behavior is inconsistent with the gradual onset expected from a distribution of single-particle tunneling energies. Instead, the observed $\sim 10~\mu$eV sharpness, limited only by measurement resolution, indicates a collective transition of the correlated electron fluid, in which chiral conduction paths reorganize globally once a critical electrochemical potential gradient is reached.

The temperature dependence provides the first key signature of many-body physics. At $T = 10$ mK and $\nu = 1.003$, transport remains fully quantized up to a sharp threshold at $V_{\mathrm{DC}} \sim 60$ mV, with zero magnetoresistance below threshold. This demonstrates that chirality and global topological order persist despite continuous local rearrangements, and that dissipation only emerges when the correlated state collapses collectively. The critical bias $V_{\mathrm{th}}$ remains nearly constant up to $\sim$100 mK, then decreases systematically at higher $T$, corresponding to an effective activation energy of only $\sim 9.5~\mu$eV. This tiny energy scale is two orders of magnitude smaller than either the cyclotron or exchange-enhanced spin gaps, ruling out single-particle activation. Instead, it reflects the collective barrier for charge equilibration\textemdash the minimal energy required to reorganize screening-induced compressible regions around disorder sites~\cite{hoai2024topological}.

This picture aligns with models of impurity screening in quantum Hall systems~\cite{Laughlin1981,cooper1993coulomb,komiyama1992inter,Electrostatics_of_edge_channels}. Localized states embedded in an incompressible background are encircled by compressible metallic rings that enable correlation-enhanced screening. In our extension, chiral currents propagate along these interfaces. As the in-plane field increases, the electrochemical gradient forces formerly isolated chiral paths to reconnect, producing new global conduction channels. At an applied bias far exceeding the disorder potential scale ($eV_{\mathrm{DC}} \gg eV_{\mathrm{dis}} \sim 0.1$ mV), the screening capacity collapses catastrophically, leading to increases in resonant conductivity. This dielectric-breakdown mechanism, driven by many-body screening failure~\cite{hoai2024topological}, contrasts sharply with the percolation-driven picture of non-interacting breakdown~\cite{nachtwei1999breakdown}.

A second fingerprint emerges in the filling-factor dependence of the threshold. As shown in Fig.~1(d), $V_{\mathrm{th}}$ exhibits an exponential sensitivity to deviations of only a few parts in $10^3$ around $\nu = 1$. In disorder-based models, the threshold would vary smoothly with filling, since the impurity landscape is fixed. By contrast, in the interaction-driven scenario, small shifts in filling renormalize the compressibility of the electron liquid, which in turn modifies the effective barrier for charge redistribution. Because tunneling and reconfiguration rates are exponentially sensitive to barrier height, modest compressibility changes are amplified into the observed exponential threshold dependence. This establishes that $V_{\mathrm{th}}(\nu)$ directly reflects the many-body stability of the correlated ferromagnetic ground state. 

\begin{figure}[h]
\vspace{0pt}
\includegraphics[width=0.49\textwidth]{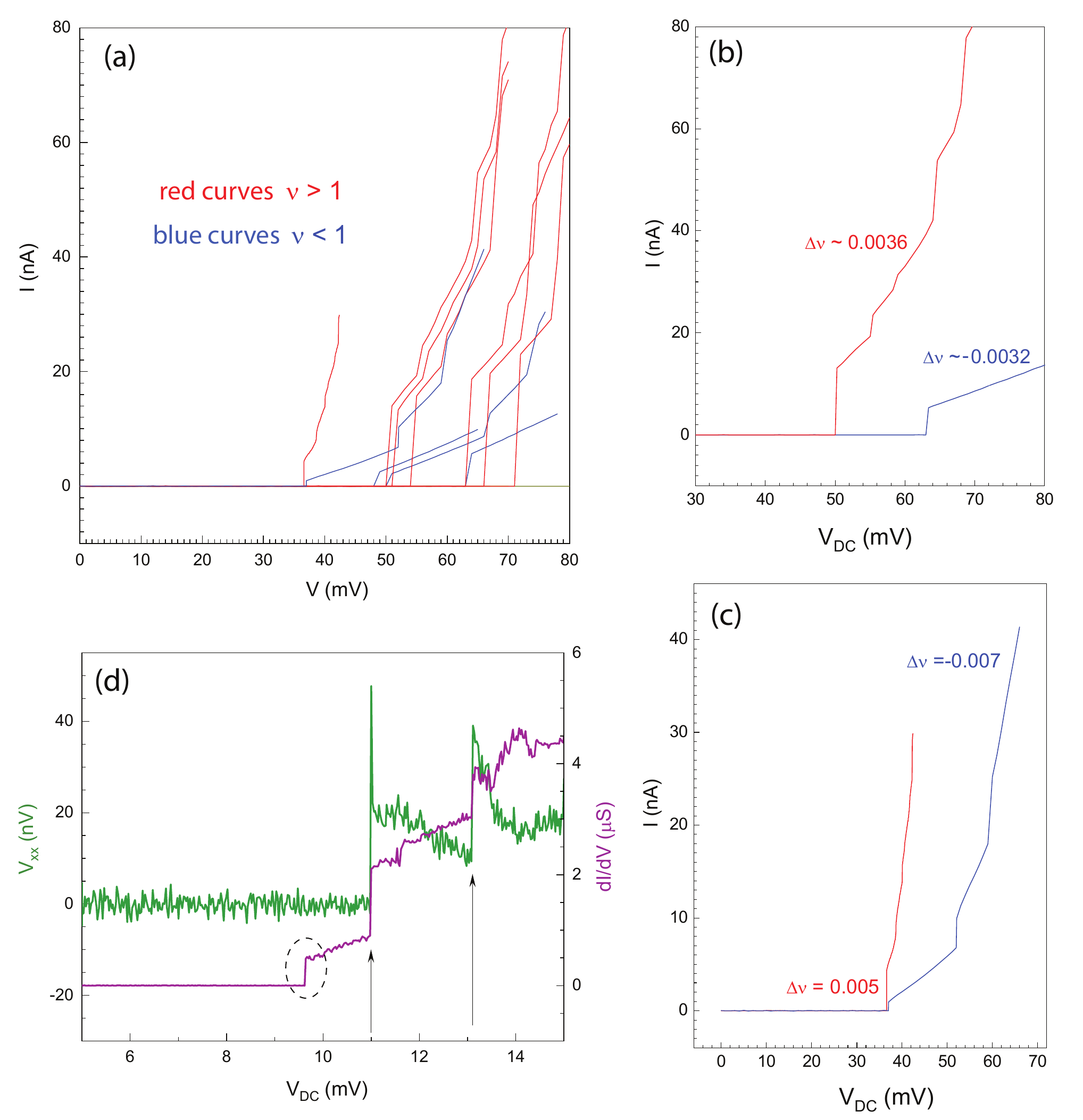} 
\vspace{-10pt}
\caption{\label{fig:Fig2}
(a) Transverse $I-V_{DC}$ characteristics for $\nu>1$ (red) and  $\nu<1$ (blue) at $T=15$~mK, showing a significantly lower dissipation for $\nu<1$. (b) Zoom-in of the dissipation onset for the curves in (a) at $\Delta\nu\sim0.0034$, highlighting the $>10$~mV difference in threshold voltage and the order-of-magnitude difference in dissipation current. (c) Comparison of $I-V_{DC}$ curves for two fillings ($\nu=$1.005 and $\nu=$0.993) where the threshold voltage is similar, yet the magnitude of the dissipative current remains starkly different. (d) Onset of dissipation in both longitudinal and transverse channels, showing a one-to-one correspondence of signal steps in real time at $\nu \sim 0.985$. The first transverse jump (dotted circle) is an artifact of the Corbino geometry and finite Hall voltage, and does not reflect the actual breakdown (see Ref.~\cite{hoai2024topological}).
}
\end{figure}

Fig.~2(a) highlights the most striking feature of our results: a pronounced asymmetry in the breakdown threshold and dissipation current between $\nu>1$ and $\nu<1$ regimes relative to the $\nu=1$ ferromagnetic IQHE. The transverse dissipation current $I$ is plotted against the applied bias voltage $V_{DC}$. The breakdown in the red curves, representing the states at $\nu>1$, exhibits a sharp increase in dissipation at a significantly lower critical voltage compared to the blue curves, which represent the states at $\nu<1$. This asymmetry sharpens at lower temperatures [Fig.~1(d)], ruling out thermal broadening. 

A detailed view of the dissipation onset is shown in Fig.~2(b) for two similar filling deviations, $\Delta\nu\sim\pm 0.0034$, above and below 1, revealing differences in the threshold voltage exceeding 10 mV and dissipation currents by approximately an order of magnitude. In addition, for two cases with similar breakdown thresholds are compared in Fig.~2(c), the filling factors differ by nearly 40\%, with stark differences in the magnitude of their dissipative current response. These behaviors highlight the role of many-body excitations, which we discuss below.

This result demonstrates that the many-body state is inherently more fragile and breaks down more readily for $\nu>1$ than for $\nu<1$. The order-of-magnitude difference in dissipation current at a given bias further underscores significant differences in the stability of the many-body state on either side of integer filling.


We point out that, using frequency-resolved detection, the longitudinal and transverse channels are cleanly separated. Fig.~4(d) show that the onset of dissipation in both directions is not only distinct but also synchronous in real time. This one-to-one correspondence demonstrates that both channels detect the same microscopic tunneling events, consistent with the collective reconfiguration of chiral current paths. The simultaneity rules out spurious effects such as contact artifacts, leakage, or edge contributions, and confirms that the breakdown asymmetry originates from intrinsic bulk processes.


\section{DISCUSSION AND SUMMARY}

Our results establish that the breakdown of the $\nu=1$ integer quantum Hall state is not a disorder-driven percolation process but an interaction-driven instability of the correlated electron liquid. The activation energy of only $\sim 10~\mu$eV is far below single-particle cyclotron or spin gaps, identifying the relevant barrier as a collective one associated with the collapse of correlation-enhanced screening around localized states~\cite{Laughlin1981,jain1988quantum,komiyama1992inter}. Breakdown occurs when the electrochemical potential gradient forces compressible regions to reorganize globally, creating new conduction channels.

A second hallmark is the exponential dependence of the breakdown threshold on filling factor. Because impurity configurations are fixed, disorder-induced percolation would only yield smooth variations. The exponential sensitivity we observe instead indicates that even small deviations from $\nu=1$ renormalize the compressibility of the incompressible background in a highly nonlinear fashion. Compressibility thus emerges as the key parameter that controls the robustness of integer quantization.

Most strikingly, the instability is strongly asymmetric across $\nu=1$. In a non-interacting Landau level the breakdown threshold would be symmetric, but in the correlated $\nu=1$ ferromagnet the situation is fundamentally different. Exchange interactions stabilize a spin-polarized ground state whose lowest-energy excitations are skyrmions ($\nu > 1$) and antiskyrmions ($\nu < 1$)~\cite{sondhi1993skyrmions,barrett1995optically,schmeller1995evidence}. These textures differ in both size and energetics: skyrmions, which spread spin distortions over many orbitals, lower their Coulomb cost and enhance compressibility, whereas antiskyrmions are more localized and less able to screen. This imbalance produces weaker reconstruction barriers for $\nu > 1$ than for $\nu < 1$, directly accounting for the observed asymmetry in breakdown thresholds. The asymmetry is therefore not an incidental detail but a fingerprint of the underlying topological spin texture.

Taken together, the ultra-low activation energy, exponential filling dependence, and asymmetric thresholds demonstrate that electron–electron interactions are indispensable even in integer topological phases. The Hall plateau remains quantized, but its stability is determined by many-body entanglement, screening, and compressibility. These findings revise the conventional view that interactions are peripheral to integer topology, showing instead that they are a defining ingredient of its robustness and pointing toward a unifying framework that places both integer and fractional states within the landscape of correlated topological matter.

In conclusion, by combining Corbino geometry with frequency-resolved transport detection, we identify a correlation-driven mechanism of breakdown in the $\nu = 1$ IQHE. The ultralow activation energy reflects a collective barrier for reorganizing chiral paths, whereas the asymmetry across $\nu = 1$ reveals the role of skyrmion versus antiskyrmion excitations. These findings demonstrate that integer topological phases cannot be fully understood without explicitly incorporating electron–electron interactions.

\section{acknowledgment}
The NSF supported this work under No.~DMR-1410302. The authors thank Professor Bertrand Halperin and Professor Joseph Avron for stimulating comments and discussions and Professor Klaus von Klitzing for raising important questions. The GaAs crystal growth work was partially funded by the Gordon and Betty Moore Foundation through Grant No.~GBMF2719 and by the National Science Foundation No.~MRSEC-DMR-0819860 at the Princeton Center for Complex Materials.
 
\newpage
\bibliographystyle{naturemag}

\bibliography{biblio}
\newpage

\end{document}